\documentstyle[iopconf1,epsf]{article}
\newcommand{\be}{\begin{equation}}
\newcommand{\ee}{\end{equation}}
\newcommand{\bear}{\begin{eqnarray}}
\newcommand{\ear}{\end{eqnarray}}

\newsavebox{\LSIM}
\sbox{\LSIM}{\raisebox{-1ex}{$\ \stackrel{\textstyle<}{\sim}\ $}}
\newcommand{\lsim}{\usebox{\LSIM}}
\newsavebox{\GSIM}
\sbox{\GSIM}{\raisebox{-1ex}{$\ \stackrel{\textstyle>}{\sim}\ $}}
\newcommand{\gsim}{\usebox{\GSIM}}

\begin{document}
\title{Phase Transitions in the Early Universe
\footnote{Talk presented at the conference ``Early Universe
and Dark Matter''(Heidelberg, July 1998)}
\protect\\
(Is there a Strongly First Order Electroweak Phase Transition?)}
\author{Michael G. Schmidt\footnote{e-mail: m.g.
schmidt@thphys.uni-heidelberg.de}}
\affil{
Institut f\"ur Theoretische Physik der Universit\"at
Heidelberg,
Philosophenweg 16, D-69120 Heidelberg}

\vspace{2cm}
\beginabstract
After some introductory remarks about the prospects of first order
phase transitions in the early universe, we discuss in some detail
the electroweak phase transition. In the standard model case a clear
picture is arising including perturbative and nonperturbative effects.
Since in this case the phase transition is not strongly first order
as needed for baryogenesis, we discuss supersymmetric variants of
the standard model, the MSSM with a light stop$_R$ and a NMSSM model
where this can be achieved. We conclude with some remarks about the
technical
procedure and about possible effects of a strongly first order electroweak
phase transition including baryogenesis.
\endabstract

\section{Introduction: Phase Transitions at High Temperature}

As also witnessed by this conference, there is a very fruitful connection
between cosmology and elementary particle physics: Accelerator
experiments and our theoretical understanding of elementary particle
processes constitute a solid basis for detailed calculations of
effects in the early universe up to temperatures $T\sim $ 100
GeV - 1 TeV of the order of the weak scale, which is the borderline
of our present understanding. On the other hand, cosmological considerations
restrict possible extensions beyond the standard model which most of
us feel to be necessary to finally obtain a theory including Planck
scale physics.

The Universe expands and cools down, and just like in
terrestrial experiments with alloys, vapor-liquid systems,
superconductivity etc. we expect phase transitions to occur
\cite{4};
relativistic kinematics would be involved and they could be very
rapid. It is a most interesting question whether such phase transitions
specific for a certain set of fields and couplings
have left traces in our universe and whether we perhaps can select
between the various ``beyond the standard'' models.

Phase transitions can be discussed in a simple way by
inspection of a temperature $T$ dependent (Ginzburg-Landau type)
effective potential in some order parameter (field) $\varphi$:
a change in the minimum of $\varphi$ due to some thermal mass term
$\sim T^2\varphi^2$ from $<\varphi>=0$ at large $T$ to
$<\varphi>\not= 0$ at $T<T_c$ leads to
a spontaneous symmetry breaking if $\varphi$ carries some charge,
and signals a phase transition.

The  standard model (SM)of elementary particle physics divides into
Quantumchromodynamics (QCD) for the interaction of colored quarks and gluons
with a scale of $\sim 100 $MeV and into the Electroweak Theory (EWT)
for flavored quarks, $W-Z$ vector bosons and the Higgs boson(s)
with a scale $\sim 100 $ GeV. The convergence of running gauge couplings and  
perhaps the need for baryon number violation point
towards some Grand Unified Theory (GUT) at a scale $\geq 10^{15}$ GeV
which should also be brought into some connection with
a model of inflation to fulfill the demands of cosmologists.
In table 1 we have listed these three gauge theories together with
the relevant symmetries and order parameters.

\bigskip
\begin{center} Table 1\end{center}
\begin{tabular}{l|c|c|c}
&QCD&Electroweak Theory&GUT/Inflation\\
\hline
&&&\\
Scale& 100 MeV&100 GeV&$10^{15..}$GeV\\
&&&\\
Symmetry& chiral symmetry&$SU_L(2)\times U_Y(1)$&$SU(5),SO(10),E_6$\\
&&&\\
Order& $<\bar\psi_q\psi_q>$&$<H>=\left(\begin{array}{c}
\varphi\\ 0\end{array}\right)/\sqrt2$&
$<X_{\rm GU-repr.}>$\\ parameter&&&$<{\rm Inflaton}>$\\
&&&\\
\end{tabular}

\medskip
There may be also some further intermediate scales connected
to $U(1)'$ \cite{A9} or $SU(2)_R$ \cite{A6} gauge interactions.

What could be the observable effects of such phase transitions? We can
think of

o \underline{P}rimordial \underline{D}ensity
\underline{F}luctuations (PDF) leading to fluctuations
in the microwave background and in the galaxy distribution

o Deformations of PDF

o Gravitational waves \cite{schw}

o Generation of the baryon asymmetry $(n_B-n_{\bar B})/n_\gamma
\sim10^{-10}$ after inflation. According to Sakharov this requires  besides baryon number violation C, CP violation and
nonequilibrium. Thus only GUT's and the electroweak theory are to
be discussed in this point.

o Large scale magnetic fields

o $(n_B-n_{\bar B})$ fluctuations.

In the following we will concentrate on the most violent
\underline{first order} PT's \cite{A2}
involving the condensation of critical
bubbles of the new phase after some supercooling- similar
to the condensation of bubbles of liquid from vapor, the
bubble expansion- and the coalescence of bubbles.

Thus we do not dwell on the also very interesting phenomenon
of defect formation \cite{A3}, \cite{A4},
\cite{A5}, in particular of cosmic strings. For this
a first order PT is not mandatory: The expanding
universe provides a particle horizon and the size of topological
defects (Kibble mechanism), and also some nonequilibrium. Of course,
quasi-stable configurations (topological configurations
\cite{A3}, Q-balls \cite{A7}, \cite{A8})
are the most interesting possibilities.

\begin{figure}[t]
\vspace{1cm}
\begin{picture}(180,100)
\put(60,-10){\epsfxsize7cm \epsffile{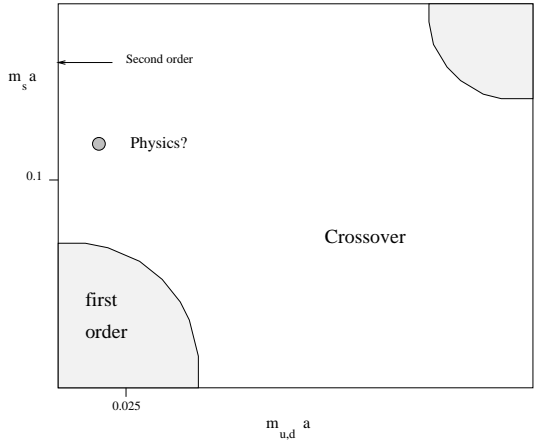}}
\end{picture}
\vspace{.4cm}
\caption{QCD phase diagram}
\label{fig1}
\end{figure}

QCD in itself is a perfect theory, its field content and
couplings are fixed. Thus the question what happens if a QCD plasma
is heated up, is well posed and should have a clear answer.
Unfortunately, this is not the case up to now. The reason is of
course that QCD is mostly nonperturbative and that this is even more
pronounced in thermal QCD. Lattice calculations are an appropriate
tool.  But particularly because fermion fields cause problems on
a lattice, many things still have to be clarified \cite{A9'}. In
fig. 1 we sketch the phase diagram \cite{A10},according 
to our present understanding,
for the case of 2+1 flavors ($m_u=m_d,m_s$). For very small
$m_u=m_d$, the critical temperature $T_c$ turns out to be of the
order of the physical strange mass $m_s$. Thus it is not clear
at all \cite{A11}
if there is a first order PT and, in case there is one, whether it
is strong. In phenomenological models (MIT bag model, chiral $\sigma$-model,
Nambu-Jona-Lasinio-type models) there seems to be a tendency to
overestimate the strength.

If there would really be a first order
PT, interesting effects could be spelled out: The speed of
sound would go to zero in course of the transition, and the primordial
density fluctuations would be deformed \cite{A12'}
with very dangerous consequences for primordial nucleosynthesis.
However peaks in the hadron-photon-lepton fluid are argued  \cite{A12'}
to be wiped out during neutrino decoupling.
Also black holes
of about a solar mass could be formed \cite{A13} but this is
controversial  \cite{A12'}. Presumably the PT
with a chemical potential (not included in fig. 2) is much more
important \cite{A14}, not so much in cosmology,
but in dense nuclear matter physics
discussed for heavy ion collisions and neutron stars.

Grand Unified Theories (GUTs) and inflation should be dealt with
together \cite{14'}
because there may be no separate GUT-PT after inflation:
The reheating temperature after inflation might be too small for
the production of GU-particles, and they might only be produced
in a preheating period \cite{A15}. Inflation in its
different scenarios, besides
being the remedy of the well-known problems of standard cosmology, is also most interesting as a source for primoridal
density fluctuations leading to structure formation. The field
content of such theories (except the postulate for a heavy neutrino?)
is rather unclear: the nature of the inflaton field,
the GU-gauge symmetry and models of a hybrid inflation. We expect
extreme supercooling for the inflation transition, an extremely
relativistic situation, perhaps no thermal state before the PT
and a complicated pre-heating process. Thus this would be anything else
but a conventional PT. If the nonequilibrium situation during and after
the phase transition (out of equilibrium decay...) is supposed to
generate the baryon asymmetry,which is the well-known textbook scenario
\cite{A16},
one has to keep in mind that the $B+L$ violating electroweak
interaction with an unsuppressed transition rate $\Gamma\sim
...(\alpha_wT)^4$ in the symmetric phase of electroweak matter
in the equilibrium period at temperatures $T\gg$ 100 GeV washes
out a previously generated baryon asymmetry if $B-L=0$. Thus
one needs a (B-L)-violating GUT gauge interaction (e.g. $SO(10)$
\cite{A17}, \cite{A17'}).

Let us then consider in detail a PT in electroweak matter. 
We will argue that it has a reliable theoretical description. As it turns
out, the standard electroweak theory does not provide 
a strongly first order PT but we will see, that it can be achieved in supersymmetric extensions of the SM.

\section{The Electroweak Phase Transition}
\setcounter{equation}{0}

The Electroweak Standard Model (SM) is so successful in
explaining high energy elementary particle processes because
the gauge couplings $g_w, g'$ of standard $SU(2)_W\times U(1)_Y$ are
small and infrared problems are tamed by the Higgs mechanism, and because
therefore perturbation theory works very well. Still there are
nonperturbative features predicted: the instanton-induced B+L-violating
interaction. However, this is an unmeasurable small effect
$\sim e^{-8\pi^2/g_w^2}$ unless it is amplified
in multi-gauge boson production \cite{B1}. This is not true anymore
for large temperatures $T>T_c$, where the B+L-violating thermal
transition rate $T\sim ...(\alpha_wT)^4$ (recently the prefactor was
discussed intensively \cite{B2}) is unsuppressed. Thus for \\
B-L=0
an existing baryon asymmetry of the universe would be erased
in the equilibrium situation before the electroweak PT. The last
chance to create the observed baryon asymmetry would then be in
course of a first order electroweak PT \cite{1,2}. In the Higgs phase with
$<\varphi>=v(T)$, the asymmetry should ``freeze out'', i.e.
the thermal sphaleron transition rate should be sufficiently
Boltzmann-suppressed $\sim e^{-const\  v(T)/T}$. Thus
one has to study the PT carefully. In particular, infrared (IR)
effects are important if the Higgs vev is reduced . They have
to be properly taken into account.

A (naive) 1-gauge boson (+ghost) loop calculation
in a Higgs field background of the $T\not=0$ effective potential
$V(\varphi^2,T)$ (free energy) is easily performed
\cite{5}.
Substitute Matsubara frequencies $2\pi nT$ in the log
\bear\label{2.1}
&&\int\frac{d^4p}{(2\pi)^4}log(p^2_0+{\vec p}\ ^2+\frac{1}{4}g^2_w\varphi^2)
\nonumber\\
&&\Rightarrow T\sum_n\int\frac{d^3p}{(2\pi)^3}log((2\pi nT)^2+{\vec p}\ ^2+
\frac{1}{4}g_w^2\varphi^2).\ear
After renormalization (as at $T=0$) the $n\not=0$ modes produce
a function $\sim T^4f\left(\frac{m^2}{T^2}\right)$
with $m^2=\frac{1}{4}g^2_w\varphi^2$ whose expansion gives the
well-known positive thermal mass term $\sim \varphi^2T^2$. The
$n=0$ mode belongs to a 3-dimensional theory without time. After
renormalization it contributes $-ET(\varphi^2)^{3/2}$
to the effective potential; this is the famous $\varphi^3$ term. This
term would make the PT first order (fig. 2), i.e. we would have
two degenerate minima of the potential at the critical temperature
$T=T_c$. However, we should not trust this result because it
completely neglects IR effects. Its gauge dependence is strongly
reduced in 2-loop order \cite{B3} but of course this does
not dispense us from the search for possible nonperturbative
effects.

\begin{figure}[t]
\vspace{1cm}
\begin{picture}(180,100)
\put(60,-10){\epsfxsize7cm \epsffile{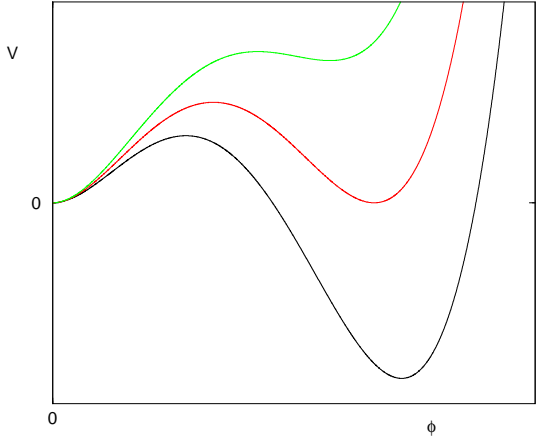}}
\end{picture}
\caption{First order phase transition effective potential}
\label{fig2}
\end{figure}

A proper way to discuss IR effects \cite{6}-\cite{9a} is
to first integrate out all nonzero Matsubara frequencies of
the theory including all fermions (with $n=1/2,..$). One thus reduces
the action to a 3-dimensional one. In a second step also the
longitudinal gauge bosons with Debye mass $m_D\sim g_wT$ can
be integrated out. Here ``integrating out'' means matching
a set of static 4-dimensional amplitudes containing the above
modes in the loop to a 3-dimensional truncated Lagrangian
for the Higgs and transversal gauge boson zero modes:
\be\label{2.2}
L_{eff}^{3-dim.}=\frac{1}{4}(F^a_{ik})^2
+(D_i\phi_3)^\dagger(D_i\phi_3)+m^2_3(T)\phi^\dagger_3
\phi_3+\lambda_3(T)(\phi_3^\dagger\phi_3)^2 .\ee
This Lagrangian then contains all the IR problems whereas
the first step can be done in (two-loop) perturbation theory
without problems. The truncation of, e.g., terms like $(\phi_3^\dagger
\phi_3)^3$ gives only an error of a few percent $(0(g^3_3)$)
\cite{7}. The theory (\ref{2.2}) is superrenormalizable. Its
dimensionful variables, the gauge coupling $g^2_3(T)$,
and $m^2_3(T), \lambda_3(T)$ can be reduced to the dimensionless
quantities
\be\label{2.3}
y=\frac{m^2_3(T)}{g^4_3(T)}\qquad x=\frac{\lambda_3}{g^2_3}\ee
$y$ is $\sim (T-T_c)$ and fixes the temperature, whereas
$x$ determines the nature of the phase transition. Of course,
$x,y$ characterize a whole class of models obtained by dimensional
reduction. The most secure way to handle (\ref{2.2}) is to put it on a
lattice \cite{9,10}. There is only one scale
$g^2_3$, we have only three dimensions, and there are no fermions. Thus
this is particularly safe. An alternative treatment would be
in the framework of Wilsonian renormalization \cite{11}. The results
of such lattice calculations \cite{9,10} are:

\begin{figure}[t]
\begin{picture}(200,165)
\put(60,-10){\epsfxsize10cm \epsffile{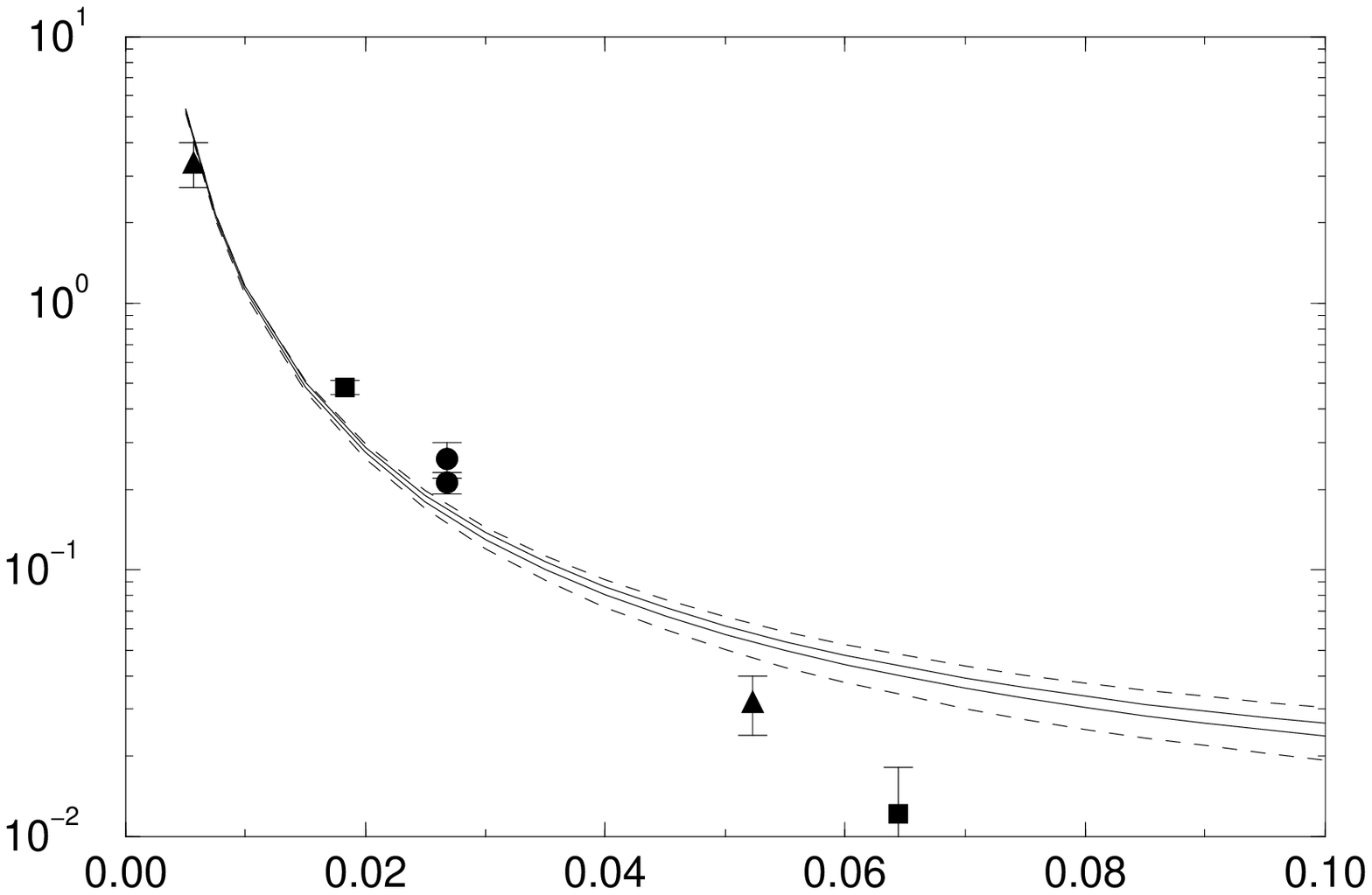}}
\put(270,0){$x\rightarrow$}
\put(60,140){$\frac{g_W^2\sigma}{(g_3^2)^3}\uparrow$}
\end{picture}
\caption{The perturbatively calculated
interface tension $\sigma$ (including $Z$-factor effect and
gauge variations) vs.~$x$ compared to lattice data
(squares, triangles and
circles) from references in the text.}
\label{fig3}
\end{figure}

(i) There is a first order PT for $x\stackrel {\scriptstyle<}{\sim}
0.11$ and there is a second-order PT at the endpoint \cite{12a}.
Above $x=0.11$ one has a crossover -- there is no PT anymore.

(ii) Comparing the 2-loop perturbative expressions obtained from
(\ref{2.2}) with lattice results, there are deviations for
$x\stackrel {\scriptstyle>}{\sim}0.05$ especially for the interface
tension (fig. 3) (from \cite{12b}, lattice points from
ref. \cite{9, 12c, 12d}).

(iii) $v(T_c)/T_c=\varphi_{min}(T_c)/T_c\stackrel {\scriptstyle>}{\sim}1$
for $x\stackrel {\scriptstyle<}{\sim}0.04$.

To protect a previously generated baryon asymmetry in a universe
with \linebreak $B-L=0$ from erasure by sphaleron transitions
$\sim \exp(-Av(T)/T)$ in a thermodynamic equilibrium
period inside the Higgs phase  one needs $v(T_c)/T_c\gsim 1$.
With $x=(1/8)m_H^2/m_W^2+c_{Pos}m^4_t/m^4_W$
where the second term alone is $>0.04$ for the observed top mass
$m_t$, this can never be achieved in the SM, independent of the
Higgs mass. Together with its CP-violating effects being smaller than
needed for an asymmetry production, this prevents the SM from explaining the
baryon asymmetry of the universe.

Lattice results give  a clear picture for the phase diagram
for the Lagrangian (\ref{2.2}). However, for some
questions like sphaleron action, shape and action of
the critical bubble -- an explicit effective (coarse-grained)
action would still be useful. It is also very important to have
some (semi)analytic picture which tells us where one can trust
perturbation theory and where not. This will be particularly true
in the case of more complicated effective actions where lattice
results may not be available. Thus we shortly discuss such a model
\cite{12}.

\begin{figure}[t]
\hspace*{-2cm}
\begin{picture}(100,100)
\put(150,0){\epsfxsize4cm \epsffile{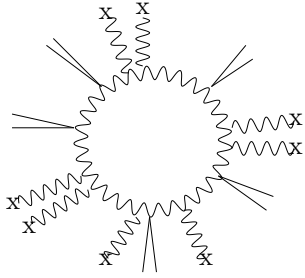}}
\put(150,26){x}
\put(156,17){x}
\put(185,98){x}
\put(199,100){x}
\put(185,5){x}
\put(223,5){x}
\put(257,58){x}
\put(257,47){x}
\end{picture}
\caption{1-loop graph contributing to the potential
$V(\varphi^2,<g_3^2F^2>)$.}
\label{fig4}
\end{figure}

In the hot symmetric phase with background $\varphi=0$ the
Lagrangian (\ref{2.2}) describes a 3-dimensional QCD-type theory
with scalar Higgs ``quarks''. Lattice calculations \cite{10} show that
indeed in this phase static ``quarks'' experience a constant
string tension which furthermore is approximately equal
to that of pure SU(2)-Yang-Mills theory. This hints to a nonperturbative
dynamics dominated by ``W-gluons''. Also a spectrum of correlation masses
of gauge-invariant $H\bar H$ bound states and of $W$-glueballs
has been calculated on the lattice \cite{13}. The former is compatible
with a linearly rising potential in a relativistic bound state
model \cite{14}
(like that of Simonov in 4-dimensional QCD \cite{15}). There is only
a small mixing with the W-glueballs \cite{13}
in agreement with the suggestion
above that we have pure ``W-gluon''dynamics.

An interesting phenomenological description of the QCD vacuum is the
``stochastic vacuum model'' of Dosch and Simonov
\cite{16,17}. Its main virtue
is that it leads very naturally to the area law of confinement.
We have applied it to the 3-dimensional theory (\ref{2.1})
with an $SU(2)_W$ gauge group. Its main ingredient is a correlated
gauge field background with a purely Gaussian correlation
\be\label{3.1}
\ll g^2_3F^a_{i\kappa}(x')F^a_{i\kappa}(x)\gg =<g^2_3
F^2>D\left(\frac{(x-x')^2}{a^2}\right)\mbox{ }.\ee
This correlator is already simplified by  choice of a
coordinate gauge and by averaging over the tensor structure.
$<g^2_3F^2>$ is the normalization by the usual local gauge
field condensate and $D$ $(D(0)=1)$ is a form factor containing the
correlation length $a$. The correlator has been
tested in 3-dimensional lattice calculations
\cite{18} and the
correlation length was obtained as $a\sim1/0.73g^2_3
\sim2/m_{glueball}$. In ref. \cite{12} we presented strong indications
that the $<g_3^2F^2>$ ground state is unstable (similar to the
Savvidy instability of QCD) for small Higgs vevs. Thus one
obtains nonperturbative effects by a fluctuating gauge field
background of the type (\ref{3.1}).

\begin{figure}[t]
\hspace*{-2cm}
\begin{picture}(200,160)
\put(-140,-20){\epsfxsize7cm \epsffile{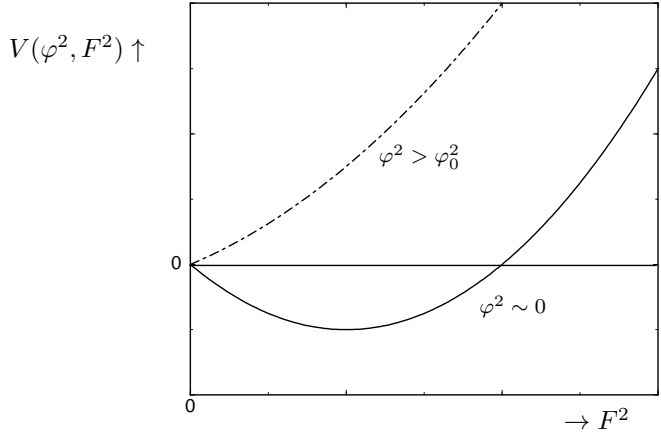}}
\put(260,-1){$\rightarrow F^2$}
\put(50,140){$V(\varphi^2,F^2)\uparrow$}
\put(228,43){\footnotesize{$\varphi^2\sim 0$}}
\put(190,100){\footnotesize{$\varphi^2>\varphi^2_0$}}
\end{picture}
\caption{Sketch of the potential $V(\varphi^2,<g_3^2F^2>)$ in
$F^2$-direction for two different values of $\varphi^2$.}
\label{fig5}
\end{figure}

\begin{figure}[p]
\hspace*{3cm}
\begin{picture}(200,180)
\put(-240,10){\epsfxsize7cm \epsffile{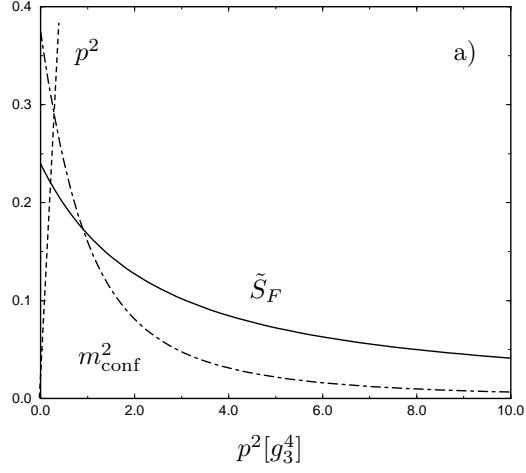}}
\put(175,170){a)}
\put(90,20){ $p^2[g_3^4]$}
\put(32,170){$p^2$}
\put(33,55){$m^2_{\rm conf}$}
\put(98,80){$\tilde{S}_F$}
\end{picture}
\caption{$m_{\rm conf}^2(p^2,m^2)$and
$\tilde{S}_F(p^2,m^2)$
in units of $(g_3^2)^2$ plotted for $m^2=0$.}
\label{fig6}
\end{figure}

\begin{figure}[p]
\hspace*{1.5cm}
\begin{picture}(200,180)
\put(-205,10){\epsfxsize7cm \epsffile{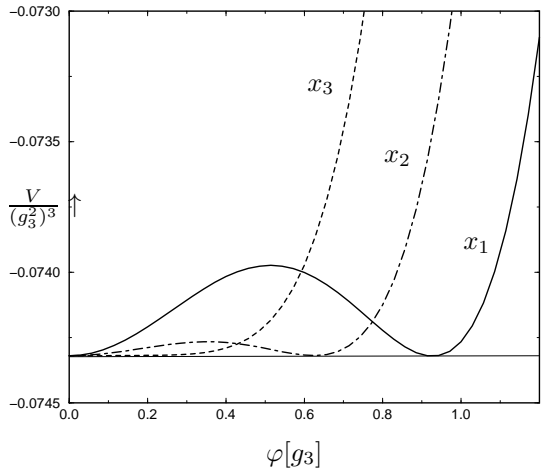}}
\put(125,20){ $\varphi[g_3]$}
\put(30,114){$\frac{V}{(g_3^2)^3}\uparrow$}
\put(144,160){$x_3$}
\put(173,133){$x_2$}
\put(203,100){$x_1$}
\end{picture}
\caption{Fading away of the first order phase transition
with increasing $x=\frac{\lambda}{g_3^2}$, where $x_1=0.06$,
$x_2=0.08$ and $x_3=0.11$.}
\label{fig7}
\end{figure}

One can estimate the effect of such a background on the W-boson
(and ghost) loop leading to the 1-loop effective potential
$V(\varphi^2,<g^2_3F^2>)$ (fig. \ref{fig4}). We found \cite{12} two
contributions to a momentum-dependent effective (``magnetic'') mass:

(i) an IR regulator mass $m^2_{conf}(p^2,\varphi^2,<g^2_3
F^2>)$ of gauge bosons and ghosts due to
the string tension (area law) which
cures the IR problems of perturbation theory.

(ii) a negative effective (mass)$^2$ for the W-bosons $-\tilde S_F
(p^2,\varphi^2,<g^2_3F^2>)$ due to spin-spin forces which becomes
important for larger $p^2$ (``paramagnetism'') and does not
spoil the nice IR properties of $m^2_{conf}$.
If we introduce these masses in the 1-loop action (gauge boson loop)
it has roughly the form
\bear\label{3.2}
V(\varphi^2,<g^2_3F^2>)&\sim&...\int\frac{d^3p}{(2\pi)^3}log[p^2
+\frac{1}{4}g^2_3\varphi^2+m^2_{conf}
(p^2,\varphi^2,<g^2_3F^2>)\nonumber\\
&&-\tilde S_F(p^2,\varphi^2,<g^2_3F^2>)] \mbox{ }.
\ear
(This has to be corrected \cite{12} for combinatorics and also has to be
renormalized). Both masses depend on $<g^2_3F^2>$. Expanding (\ref{3.2})
up to first order in
$<g^2_3F^2>$, the spin-spin force in
$-\tilde S_F$ produces the well-known negative
$F^2$-term destabilizing the $F^2=0$ vacuum. Adding the tree
$\frac{1}{4}F^2$ we can obtain an effective potential sketched in
fig. \ref{fig5} stabilized  at some value $F^2\not=0$ by confinement forces.
This is a 1-loop calculation and the masses
$m^2_{conf}$ and $-\tilde S_F$ are
determined only roughly (in lack of lattice data support). Thus we have
only a qualitative picture. To proceed, we fixed $<g^2_3F^2>$ at
the minimum by a relation to the lattice string tension.

Fig. \ref{fig6} \cite{12} shows the qualitative form of
$m^2_{conf}(p^2,\varphi^2)$ and
of $\tilde{S}_F(p^2,\varphi^2)$.  Fig. {\ref{fig7}
\cite{12} presents the new 1-loop
potential at the critical temperature at various $x$-values,
and one can see the first order PT fading away. One can
also evaluate the interface tension (table) and roughly determine
the endpoint of the PT by postulating that the effective $\varphi^2$
and $\varphi^4$ vanish at this (conformal) point with a
second-order PT.

\vspace{0.5cm}
\centerline{
\begin{tabular}{|c||c|c|} \hline
$x$ & $\sigma$ & $\sigma_{perturbative}$ \\ \hline \hline
0.06 & 0.016 & 0.013 \\ \hline
0.08 & 0.004 & 0.007 \\ \hline
0.11 & 0 & 0.004 \\ \hline
\end{tabular} }
\vspace{0.5cm}

We should stress again that this picture of nonperturbative
effects is not really quantitative, in particular because
2-loop calculations  in a correlated gauge-field background are
(too) difficult. Still we might get an indication in which
direction nonperturbative contributions go.
In this context also the work \cite{gleis} on subcritical
bubbles should be mentioned. If the crossover can be
described in this picture is an interesting question.

\section{ The MSSM with a ``light'' stop}
\setcounter{equation}{0}

Searching for modifications of the electroweak theory in
order to obtain a strongly first order PT, one
faces  the by now sufficiently known situation
that the success of the standard model is both blessing \underbar{and}
burden. We do not have experimental hints which way to go.
Supersymmetric theories have the well-known theoretical
advantages. From a practical point of view, all one needs
for a strongly first order PT is the strengthening of the
``$\varphi^3$''-term in the effective potential due to bosonic
exchange in the loop. Thus one needs further bosons with a strong
coupling to the Higgs. SUSY models have a host of new bosons
in the superpartner sector. In particular, the $s$-top particles
have a particularly strong Yukawa coupling $h_t$ if the Higgs vev $<v_2>$
of the Higgs coupling to the top
$(m_{top}=h_t<v_2>)$ is not very large, i.e., if
$\tan\beta=v_2/v_1$ is not large. The superpartner of the right-handed
top, the $stop_R$, does not have
$SU(2)_W$ interactions, and thus is particularly flexible in its
allowed mass (no $\rho$-parameter problem). As proposed in ref. \cite{19,20},
its exchange (fig. \ref{fig8})
\begin{figure}[t]
\hspace*{-2cm}
\begin{picture}(100,100)
\put(150,0){\epsfxsize4cm \epsffile{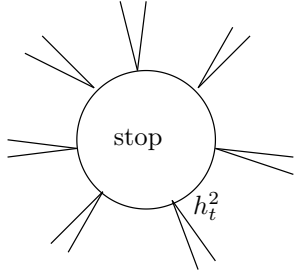}}
\put(193,50){stop}
\put(223,24){$h_t^2$}
\end{picture}
\caption{1-loop stop contribution to the effective potential.}
\label{fig8}
\end{figure}
can enhance the PT significantly if its
mass $m^2_3$ in the symmetric phase (including $T^2$-plasma mass)
is small:
\be\label{4.1}
m^2_3=m^2_0+cT^2  ,\ee
where $m^2_0$ is the SUSY-breaking scalar mass of the $stop_R$.
The $T=0$ mass of the $stop_R$ is
\be\label{4.2}
m^2_{\tilde t}=m^2_0+m^2_t\ee
and is not much larger than the top mass for small positive
$m^2_0$. A nonuniversal SUSY mass breaking
at the GUT scale might be necessary for very small $m_0^2$  although the
stop mass$^2$ is naturally lowered by renormalization flow.

If the $stop_R$ and one heavy combination of Higgses
is integrated out, one is led again to a Lagrangian
of the form (\ref{2.1}), but now with an $x=\lambda_3/g^2_3$ value
much smaller than in the SM (being bosonic, the stop contributes opposite to
the top!)
allowing for $m_H\lsim 75$ GeV for a strongly
first order PT with $v(T_c)/T_c\geq1$ \cite{22}-\cite{25}.

One can also ask \cite{19} for $stop_R$ masses smaller than the top mass
taking $m^2_0=-\tilde m^2_0$ negative in
(\ref{4.1}), (\ref{4.2}). Then the  $stop_R$ than should not be fully
(also zero modes) integrated out, but kept in the effective 3-dimensional
theory together with the light Higgs fields. If one
assumes that the CP-odd Higgs $A_0$ meson surviving spontaneous breaking
is rather heavy $(\gsim 300$ GeV), there is a heavy Higgs sector to
be integrated out, and just as above one Higgs field remains. Thus we have
to consider a Lagrangian \cite{23}
\bear\label{4.3}
L^{3-dim}_{eff}&=&L^{3-dim}_{eff}(Higgs)\nonumber\\
&&+\frac{1}{4}G^A_{ij}G^A_{ij}+(D^s_iU)^+(D_i^sU)+
m_{U_3}^2U^+U\nonumber\\
&&+\lambda_{U_3}(U^+U)^2+\gamma_3(\phi^+_3\phi_3)(U^+U) \mbox{ }.\ear
The $T$-dependent parameters are obtained by integrating
out all non-zero modes and all heavy particles like in (\ref{2.2}),
which is the first part of the Lagrangian (\ref{4.3}). Thus one
has to specify the field content and the SUSY-breaking
parameters of the model. The simplest choice is the minimal
supersymmetric standard model (MSSM) \cite{21}
without universality for the top
scalar SUSY-breaking masses. The partner of the left-handed top
with a SUSY-breaking mass $m^2_Q$ should be heavy in order not to
contribute too much to $\Delta\rho$.

\begin{figure}[p]
\hspace*{-2cm}
\begin{picture}(200,170)
\put(95,-10){\epsfxsize7cm \epsffile{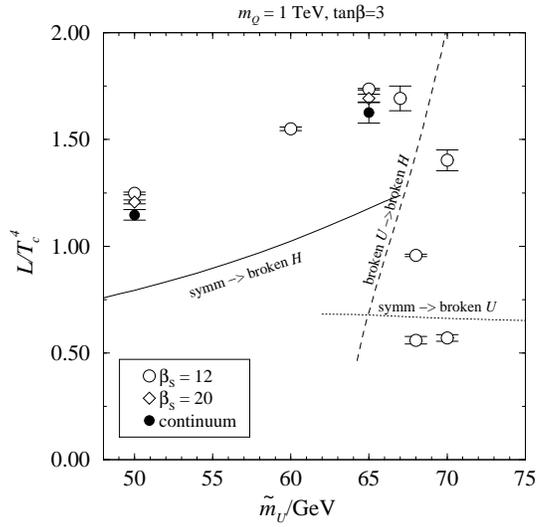}}
\end{picture}
\caption{Latent heat at $T_c$ as a function of mass parameter
$\tilde{m}_U$ calculated on the lattice [54] compared to
the analytic results [52].}
\label{fig9}
\end{figure}

\begin{figure}[p]
\hspace*{-2cm}
\begin{picture}(200,120)
\put(-135,-10){\epsfxsize7cm \epsffile{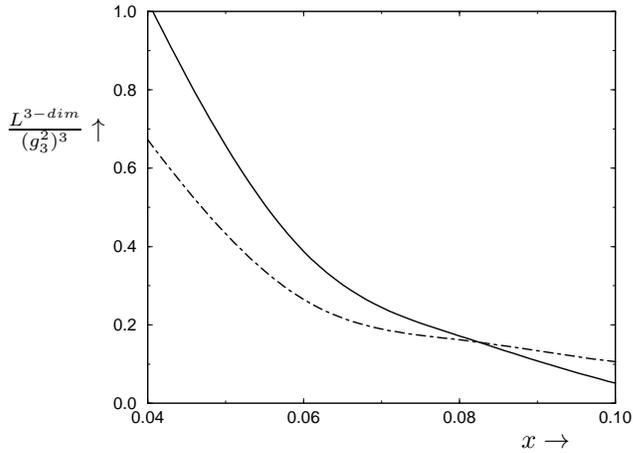}}
\put(265,5){$x\rightarrow$}
\put(70,123){$\frac{L^{3-dim}}{(g_3^2)^3}\uparrow$}
\end{picture}
\caption{Three dimensional latent heat as a function  of $x$
calculated from the potential (2.5) (full line) compared to
the result of ordinary 1-loop perturbation theory
(dashed-dotted line).}
\label{fig10}
\end{figure}

Two-loop calculations with (\ref{4.3}) have shown that one can
indeed obtain \linebreak $v(T_c)/T_c\gsim1$ even for lightest Higgs masses
as big as 105 GeV \cite{26}. The parameter
space is enlarged \cite{27}
if one allows for $stop_R$-$stop_L$ mixing with a parameter
$\tilde A_t=A_t+\mu ...$ . (Both parameters $\mu$ and
$A_t$ are important in the discussion of CP-violations
in the wall.) Ref. \cite{27} uses an improved 4-dimensional
one-loop effective potential at high temperatures and still
agrees well with the special case considered in \cite{26}.

For large enough negative $m^2_U=-\tilde m^2_U$ one even
obtains \cite{26,27} a two-stage phase transition with an intermediate
stop condensate $<U^+U>$. This is only acceptable if the
transition rate which is rapidly decreasing  with increasing
$\tilde m^2_U$, still allows to return from the stop phase
to the Higgs phase. In the former phase one has a situation
analogous to the Higgs phase, in particular,there are
massive $SU(3)$ gauge bosons.

Recent lattice calculations confirm the perturbative results \cite{26}
surprisingly well \cite{28} (fig. \ref{fig9}) --
though there are also significant
deviations. In particular the PT turned out to be more strongly
first order -- the latent heat and $v(T_c)/T_c$ are larger
than in the perturbative result. We can understand this effect
qualitatively with our model for nonperturbative contributions:
The effective $x$-value in the Higgs part of (\ref{4.3}) is much
smaller than in the standard model and for these values (fig. \ref{fig10})
the latent heat and $v(T_c)/T_c$ are both increased compared to
pure perturbation theory. The important additional graphs coming
from Lagrangian (\ref{4.3}) mostly involve $SU(3)$ gluons
and the $stop_R$ both of which do not have $SU(2)_W$ interaction,
and hence also no nonperturbative effects on
this scale.

\section{NMSSM with a strongly
first order phase transition [55]}
\setcounter{equation}{0}

In the effective electroweak potential near the critical
temperature a term of type $-\varphi^3$ triggers a first order
PT. Up to now we discussed the generation of such
terms in 1-loop order of perturbation theory. There is also
the possibility to obtain it already on the tree level. An
$SU(2)_W$-invariant third-order polynomial term in the
potential cannot just contain the Higgs(es). Thus one has
to enlarge the field content of the SM and also of the MSSM
in the case of a supersymmetric theory. The simplest extension
of the MSSM, the ``next to minimal model'' NMSSM \cite{29,29a}, contains
a further superfield $S$, which is a gauge singlet,
in an additional piece of the superpotential
\be\label{5.1}
g^S=\lambda SH_1H_2-\frac{k}{3} S^3.\ee
The soft SUSY breaking term
\be\label{5.2}
V^S=A_\lambda \lambda SH_1H_2-\frac{k}{3}A_kS^3\ee
has the desired ``$\varphi^3$'' form \cite{30} if $S$ can be treated similarly to the $H_i$. The superpotential (\ref{5.1})
has the virtue of avoiding the $\mu$-term $g^\mu=\mu H_1H_2$
with its fine-tuning problem because
this term automatically arises after the singlet field acquires a vev. However,
because of its $Z_3$ symmetry it suffers from  the well-known
domain wall problem \cite{31}\cite{32b}.
It turns out that the NMSSM with just
(\ref{5.1}) and (\ref{5.2}) besides having the domain wall problem
also is unable to produce a phase transition in $<S>$ and
$<H>$ simultaneously, which requires $<S>$ and $<H>$ to be
of the same order of magnitude. With a very large $<S>$
one would first obtain a PT in $<S>$ and afterwards the ordinary MSSM
PT in some Higgs field combination, which is not what we want.
We thus as in ref. \cite{32} choose the superpotential
\be\label{5.3}
g=g^S+\mu H_1H_2-rS.\ee
Unlike in ref. \cite{34} more than a decade  ago, we now  keep the full
parameter space of the model only restricted by universal
SUSY breaking at the GUT scale. In the latter we differ
from ref. \cite{32} where the parameters were fixed at the
electroweak scale without such a criterion. Besides the well-known
gauge couplings in the $D$-terms we then have the parameters
$\lambda, k,\mu,r$ in the superpotential, and-for the SUSY breaking
a universal scalar mass squared $m^2_0$ , a common gaugino mass $M_0$,
as well as an analytic mass  term $B_0$ for the Higgses and a
universal trilinear scalar coup- \linebreak ling $A_0$ corresponding to the
second and third power terms in the superpotential, respectively.

Besides the tree potential and 1-loop Coleman-Weinberg corrections
we include 1-loop plasma masses for the $H_i$ and $S$ fields and
the 1-loop ``$\varphi^3$''
terms discussed in previous chapters which, however, now in
general are small compared to the tree term (\ref{5.2}). The most
important finite temperature contributions come from the top quark
and the gauge bosons, but in some parts of the parameter space
the stops, charginos and neutralinos may become rather light
and therefore are also included in the effective potential $V_T(H_1,H_2,S)$.

Having at hand the potential we are  interested in, a rather natural
procedure would be as follows: (Randomly) choose a set of the GUT scale
parameters listed above. Then use the (1-loop) renormalization
group equations \cite{35} to evolve the parameters down to the
weak scale and minimize the T=0 effective potential in order to
study the electroweak symmetry breaking. Of course,in order to reproduce
the physical Z-boson mass $M_Z$, a rescaling of all the (unknown)
dimensionful parameters is necessary. But
after this rescaling in almost all cases
there appear some unobserved light
particles in the spectrum, so one has to try the
next set of parameters and this whole ``shot-gun'' procedure is
very inefficient.

Instead, we fix the T=0 electroweak minimum determined by $M_Z$,
$\tan\beta=v_2/v_1$ and  $<S>$ in addition to the parameters
$\lambda,k,m_0^2,M_0$,$A_0$ while $\mu$,$r$,$B_0$
remain unspecified. The important thing is that the latter
do not enter the 1-loop renormali- \linebreak zation group equations for
$\lambda$, $k$ and the
soft parameters except $B$.Thus we can calculate all
parameters of the effective potential at the weak scale except
$\mu$, $r$ and $B$ which we determine by applying the
minimization conditions
\[\frac{\partial V_{T=0}(H_1,H_2,S)}{\partial H_i}=0 \quad, \quad\quad\quad
\frac{\partial V_{T=0}(H_1,H_2,S)}{\partial S}=0 \quad.\]
Because of the complicated 1-loop
corrections these equations cannot be solved analytically,but an
iterative numerical solution taking the tree level solution as
starting values is possible.  Of course, whether the postulated
minimum $(M_Z,\tan\beta,<S>)$ is indeed the global minimum has
to be checked explicitly and constrains the para- \linebreak
meter space of the model.
Using this procedure we are left with the seven parameters\footnote
{Additionally, we require the top quark mass $M_{top}=175$ GeV
which allows us to fix the top Yukawa coupling as a function of
$\tan\beta$. All the other Yukawa couplings are neglected which
is only justified in the regime $\tan\beta\lsim 10$.}
\[\tan\beta,<S>,\lambda,k,m_0^2,M_0,A_0\]
which still contain a lot of freedom. Fortunately, not all parameters
are equally important with respect to the strength of the PT:
Of most interest are the gaugino mass $M_0$ and the trilinear
scalar coupling $A_0$, as they determine the coefficients
$A_{\lambda}$ and $A_k$ of the  ``$\varphi^3$''-terms in eq.
(\ref{5.2}). Therefore we will study the plane of these parameters
while keeping the others fixed. To maximize the lightest CP-even
Higgs mass $M_h$ $\tan\beta$ should be taken large while
$\lambda$ should be kept small. As stated above, a strong
PT can only be expected, if $<S>\sim M_Z$ which requires
$k$ to be not too small because of $<S>\sim \frac{A_k}{k}$.
The remaining parameter $m_0^2$ only influences the masses of
the additional Higgs bosons which we have chosen to be  heavy.

\begin{figure}[t]
\hspace*{-2cm}
\begin{picture}(180,220)
\put(-185,-20){\epsfxsize8cm \epsffile{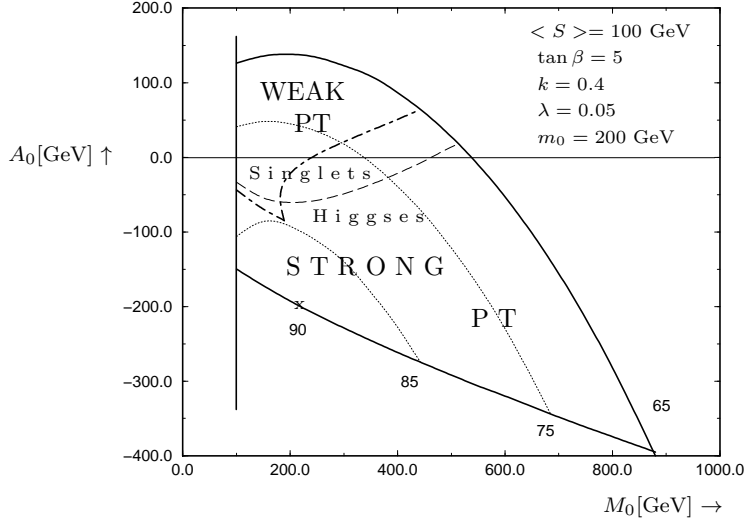}}
\put(270,-5){\footnotesize{$M_0$[GeV] $\rightarrow$}}
\put(45,128){\footnotesize{$A_0$[GeV] $\uparrow$}}
\put(153,72){\scriptsize{x}}
\put(242,175){\scriptsize{$<S>=100$ GeV}}
\put(245,165){\scriptsize{$\tan\beta=5$}}
\put(245,155){\scriptsize{$k=0.4$}}
\put(245,145){\scriptsize{$\lambda=0.05$}}
\put(245,135){\scriptsize{$m_0=200$} GeV}
\put(140,151){WEAK}
\put(153,139){PT}
\put(150,85){S T R O N G}
\put(220,65){P T}
\put(136,121){\scriptsize{S i n g l e t s}}
\put(160,105){\scriptsize{H i g g s e s}}
\end{picture}
\caption{Scan of the $M_0$--$A_0$ plane where the remaining
parameters are fixed. The full line surrounds the phenomenologically
viable part of the parameter space. The dotted lines are curves of
constant lightest Higgs mass (75 and 85 GeV). The dashed line
indicates the region where the lightest Higgs is predominantly
a singlet. The dashed-dotted line separates the regions of strong
($v_c/T_c\gsim 1$) and weak PT.}
\label{fig11}
\end{figure}

An example of a scan in the $M_0$--$A_0$ plane is shown
in fig. (\ref{fig11}) where we fixed the remaining parameters
according to the remarks before as $<S>$=100 GeV,
$\tan\beta$=5, $\lambda$=0.05, $k=$0.4 and $m_0$=200 GeV.
There are several constraints on the parameter space: First of all,
the minimum postulated in the elimination procedure discussed
above has to be the global minimum which leads to the lower
bound on $A_0$ in fig. (\ref{fig11}). To prevent the appearance of
a chargino with mass smaller than 80 GeV the gaugino mass
$M_0$ has to be larger than 100 GeV corresponding to the
vertical line in the plot. Finally, we require the lightest Higgs
mass $M_h$ to be larger than 65 GeV which leads to the
upper bound on $A_0$ in fig. (\ref{fig11})\footnote{Note that this
also implies an upper bound on the gaugino mass depending
on the remaining parameters.}. Compared with the
current LEP data on SM-like Higgs bosons this may seem
to be a rather low value, but one has to keep in mind that
the lightest ``Higgs'' state in this model always has some
singlet component which even dominates in the region above
the dashed line. Therefore the experimental constraints
on $M_h$ are somewhat relaxed.

In order to investigate the strength of the PT we determine
the critical temperature $T_c$
at which there exist two degenerate minima in $V_T(H_1,H_2,S)$,
a broken minimum with $<H_i>\neq 0$ and a symmetric one
with $<H_i>=0$\footnote{The singlet vev is different from zero
even in the symmetric minimum.}. For the previously discussed
set of parameters the results are  summarized in fig. (\ref{fig11}).
There the dashed-dotted line separates the region with a
weak PT from the region where the baryon number washout
criterion $v_c/T_c\gsim 1$ is fulfilled. One clearly sees that
{\em most} of the parameter space is indeed compatible with
electroweak baryogenesis. Interestingly enough, the region
in which the Higgs mass is maximized ($M_h\sim 90$ GeV)
is not excluded. Let us again stress that the situation
drastically changes if we increase the singlet vev
to e.~g.~$<S>=300$ GeV while decreasing $k$ in order to
obtain similar values of $M_h$. Then only a small range
of values of $A_0$ just above its lower bound allows
a strong PT and most of the parameter space leads
to erasure of the baryon asymmetry.

In the previous example the maximal value of the Higgs mass is 90 GeV
but one can reach much higher values. By choosing $\tan\beta$=10,
$M_h$=100 GeV can be obtained and still $v_c/T_c\gsim 1$ can be
fulfilled. Increasing the singlet vev to e.~g.~$<S>=250$ GeV allows
the even larger value of $M_h$=115 GeV without violating the washout
criterion. But with larger $<S>$ the amount of fine-tuning
of $A_0$ increases and there is the danger of metastability since
the PT requires thermal tunneling over a rather high tree barrier.

\section{Discussion}

Having obtained some variant of the electroweak SM with a strongly
first order PT, the way is free for the discussion of baryogenesis,
of a lot of questions both on the conceptual,
and on the technical side \cite{45}-\cite{44}.

\noindent(i) The procedure of the PT can be worked out in
detail:
\begin{itemize}

\item  First one has to find the wall profile of the critical bubble
(and later on of the stationary expanding bubble) in general in
multidimensional field space ($H_{1,2}, S, stop,$, CP-violating
angles...). This is a very demanding numerical problem which
has only  been attacked recently. \cite{77'}

\item  Given the wall profile, one can calculate the action, and one can
calculate the transition rate using Langers's formula, discuss
supercooling, and obtain the nucleation temperature
(1 bubble/universe).

\item  The interaction of the bubble wall with the hot plasma constitutes
friction; this determines the stationary velocity of the wall in the
heat bath \cite{45}-\cite{47}. Deflagration with velocity $v_B$ smaller
than the velocity of sound in the plasma seems to be favored. The
particle mean free path usually turns out to be smaller than the
thickness of the wall (``thick wall'').

\item  In front of the proceeding wall there is thermodynamic nonequilibrium
and transport.
Hydrodynamics and Boltzmann equations come into play.

\item  After some time many expanding bubbles have formed and collide.
This finally leads to the new (Higgs) phase in the whole space
and to some reheating. This is all beset by technical problems;
but there are interesting proposals: $v(T)/T$ might be
lowered in the colliding bubbles \cite{69'},
(seed) magnetic fields may
be formed via turbulence \cite{59,61} or by the Kibble mechanism
\cite{58, 62}. The scale
of the magnetic fields seems to be too small to explain the
observed magnetic fields, but there may be some enhancement.

\end{itemize}

\medskip
\noindent (ii)
If for some reason fluctuating primordial hypercolor magnetic
fields $(Uy(1))$ are produced much before the electroweak PT via the
chiral anomaly they could produce spatial fluctuations of
$n_B-n_{\bar B}$ \cite{63}. These could be frozen in a first order
electroweak PT and they would have effects on the early
nucleosynthesis \cite{60}. Primordial $Y$-magnetic fields also could
strengthen the electroweak PT \cite{64}.

\medskip
\noindent (iii)
Last not least baryogenesis- the creation of a baryon-antibaryon
asymmetry- can be discussed very concretely, favorably in
some version of the charge transport mechanism \cite{51}:the scattering of
charginos, neutralinos, stops at  the bubble wall creates
some chiral current if CP is violated in the bubble wall
\cite{48}-\cite{50}. This
is then transformed into a baryon asymmetry by the B+L violating
``hot'' sphaleron interaction in the hot phase. CP could be
violated explicitly (in the MSSM by $A_t,\mu$) or spontaneously.
If, in the latter case, this happens only for the temperatures of
the PT, one does not have any problems with EDM-bounds \cite{50}.

In recent discussions \cite{52}-\cite{57}
it is stressed that one should deal with
(thick) wall scattering and diffusion simultaneously and that one
should perhaps use quantum Boltzmann equations. This is a demanding
program without agreement on most of the technicalities.

In conclusion one can say that for a first order electroweak
PT we have an intriguing interplay of cosmology and elementary
particle physics, and of non-equilibrium thermodynamics. Variants
of the electroweak SM like the MSSM with a ``light'' stop and
NMSSM models with $\mu\not=0$ can give a strongly first order
PT even for smallest Higgs masses of 100 GeV and perhaps even higher.
A mixture of perturbation theory, lattice and (semi)analytic
methods allows to produce reliable results
for the PT. Fortunately perturbative results are more reliable
for stronger first order PTs which are the most interesting ones.

\section*{Acknowledgments}
I would like to thank D. B\"odeker, S. J. Huber, P. John,
M. Laine, A. Laser, M. Reuter for enjoyable collaboration
on various topics of this talk and also H. G. Dosch, O. Philipsen,
Ch. Wetterich for useful discussions. This work was supported
in part by the TMR network Finite Temperature Phase Transitions
in Particle Physics, EU contract no. ERBFMRXCT97-0122.

\end{document}